\documentclass[twocolumn,prb,color,superscriptaddress,psfig,showpacs,amsmath,amssymb]{revtex4}

\usepackage{epsf}

\usepackage{natbib}
\setcitestyle{square,numbers}

\usepackage{tabularx} 
\usepackage{graphicx} 
\usepackage{epstopdf}
\usepackage{epsf}
\usepackage{dcolumn}
\usepackage{bm}
\usepackage{color, xcolor}

\usepackage{multirow}

\usepackage[normalem]{ulem}

\usepackage{color,soul}
\setul{0.5ex}{0.5ex}
\setulcolor{red}

\begin{document}

\title{Raman study of the structural transition in LiVO$_2$}

\author{Yuri S. Ponosov}
\affiliation{Institute of Metal Physics, Ural Branch of the Russian Academy of Sciences, Ekaterinburg 620137, Russia}

\author{Evgenia V. Komleva}
\affiliation{Institute of Metal Physics, Ural Branch of the Russian Academy of Sciences, Ekaterinburg 620137, Russia}
\affiliation{Department of theoretical physics and applied mathematics, Ural Federal University, Ekaterinburg 620002, Russia}

\author{Elizaveta A. Pankrushina}
\affiliation{Laboratory of Arctic Mineralogy and Material Sciences, Kola Science Centre,
Russian Academy of Sciences, Apatity 184209, Russia}

\author{Haohang Xu}
\affiliation{School of Physics, Harbin Institute of Technology, Harbin 150001, China}

\author{Yu Sui}
\affiliation{School of Physics, Harbin Institute of Technology, Harbin 150001, China}
\affiliation{Laboratory for Space Environment and Physical Sciences, Harbin Institute of Technology, Harbin 150001, China}

\author{Sergey V. Streltsov}
\affiliation{Institute of Metal Physics, Ural Branch of the Russian Academy of Sciences, Ekaterinburg 620137, Russia}
\affiliation{Department of theoretical physics and applied mathematics, Ural Federal University, Ekaterinburg 620002, Russia}

\date{\today}

\begin{abstract}
The results of polarization-dependent Raman spectroscopy of single-crystalline LiVO$_2$ exhibiting transition to a diamagnetic state below $T_c \sim $500K are reported. Our measurements clearly detect additional peaks in the low-temperature phase, which disappear nearly completely when heated above $T_c$. Proposed $\sqrt{3}$a × $\sqrt{3}$a lattice reconstruction explains these new Raman peaks by the Brillouin zone folding. The experiment, on the one hand, confirms that the symmetry of the non-magnetic phase is not lower than trigonal, but, on another hand, our thermal cycling study suggests possible stacking faults. This agrees with results of the density functional theory calculations, which show that the energy difference between different types of stacking does not exceed 1 K per formula unit.

\end{abstract}

\maketitle

\section{Introduction}\label{intro}
LiVO$_2$ belongs to class of materials with a strong interplay of orbital, lattice, and spin degrees of freedom~\cite{Khomskii2021}. It crystallizes in a rhombohedral structure with $R\bar 3m$ space group above 500 K \cite{Kobayashi1969}. Structurally, it consists of alternating Li/V and O layers and within each layer the V ions form a lattice of triangles, which are equilateral in the high-temperature phase. These ions are octahedrally coordinated and therefore V 3$d$ states are split onto doubly degenerate $e_g$ and triply (nearly) degenerate $t_{2g}$ levels, due to a ligand field of approximate $O_h$ symmetry.

Almost seventy years ago this material was found to undergo a first-order structural transition at $T_c\sim 500$ K~\cite{Bond}. This transition is accompanied by drastic changes of magnetic properties. While magnetic susceptibility follows Curie-Weiss law at high temperatures, the low-temperature phase turned out to be non-magnetic~\cite{Kobayashi1969}. Similar transitions have been observed in LiVS$_2$\cite{Katayama2009}, Na$_2$Ti$_3$Cl$_8$~\cite{Hanni2017,Kelly2019,Khomskii2021-NaTiCl}, CsW$_2$O$_6$~\cite{Hirai2013,streltsov2016,Okamoto2020,Nakamura2022} 
and some other materials. Several possible mechanisms have been proposed to explain the phase transition observed in LiVO$_2$. Goodenough argued that the in-plane V-V separation in this material is near a critical distance $R_c$ that corresponds to the situation, when the bandwidth $W$ is approximately equal to the on-site Hubbard repulsion $U$. According to Goodenough's phenomenological theory~\cite{Good} there develops an electronic instability that is manifested by the formation of V$_3$ trimers below $T_c$~\cite{Good}. An alternative scenario for the phase transition in LiVO$_2$ involves an orbital ordering of the localized V $3d^2$ electrons~\cite{Pen,Pen2}. The orbital degrees of freedom suppress frustration on triangular lattice formed by V ions. Moreover, one might also think of the orbitally driven Peierls transition~\cite{Khomskii2021}. All proposed mechanisms assume formation of a spin-singlet state in the low-temperature phase due to formation of vanadium trimers.

Indeed, superlattice reflections tentatively related to the trimerization were observed by X-ray diffraction of powder samples~\cite{Car,Onoda,Kojima2019,Gau,Pour}. The $P31m$ space group was considered as a reasonable choice for the low-temperature phase~\cite{Car}. Superstructural spots appearing at (1/3, 1/3, 0) in the diffraction patterns of LiVO$_2$ single crystals were observed in ~\cite{Hew,Tak,Im,Im1}. Structural changes at 500K transition were later seen by transmission electron microscopy (TEM) experiments performed on LiVO$_2$ single crystals~\cite{Tian}. Based on these experiments various space groups were proposed for the trimerized phase $P31m$\cite{Car}, $P3_1 12$, $P3_2 12$ ~\cite{Tak,Im,Im1}, $P6_2 22$, $P6_4 22$~\cite{Hew}. In addition, it was noted that the configuration of trimers in different layers can be different, even leading to their completely disordered state along the $c$ axis, and in ~\cite{Tak,Im,Im1} a structure (including 6 layers of vanadium) with doubling of the $c$ axis was suggested. 

\begin{figure}
\centering
\includegraphics[width=1\columnwidth]{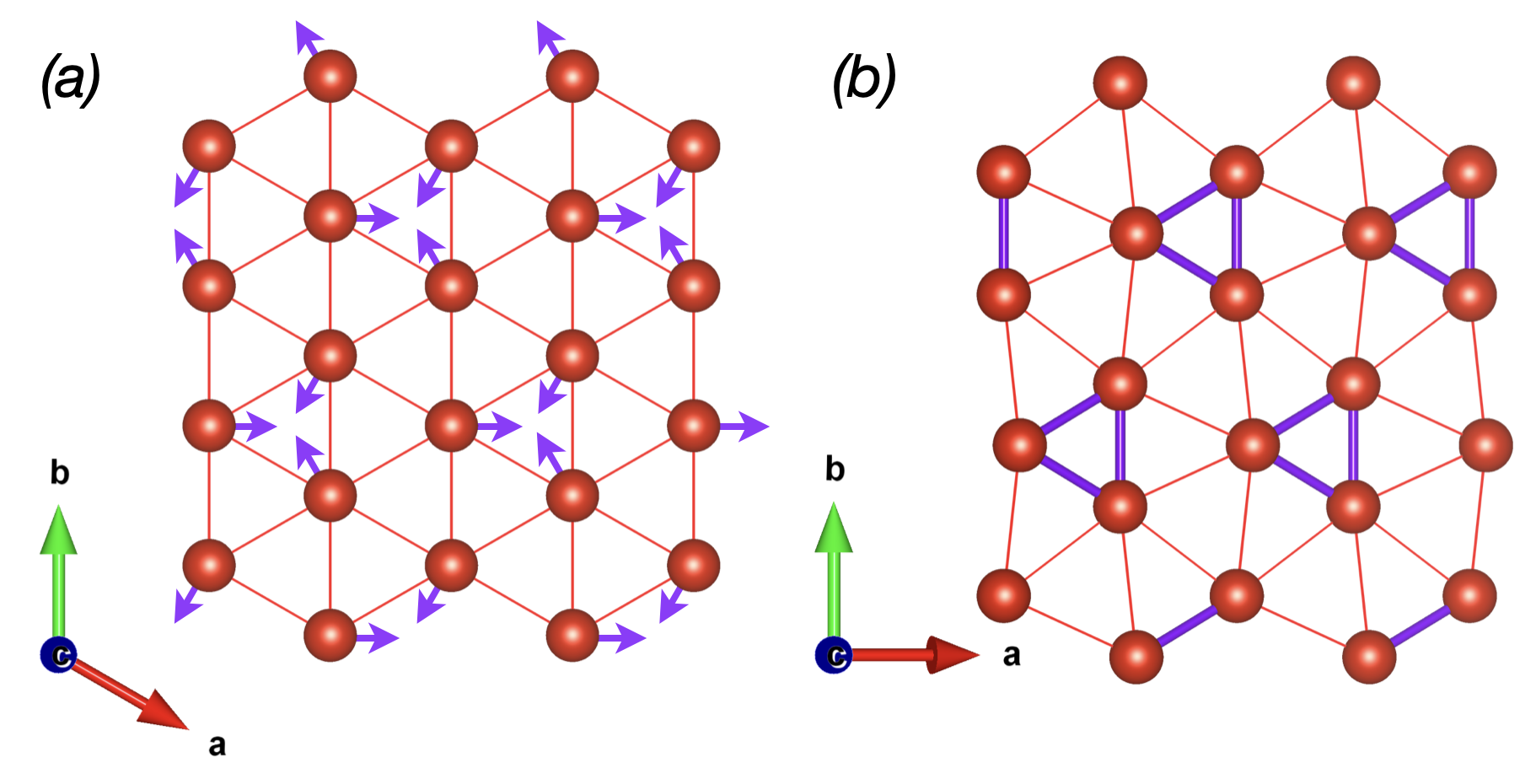}
\caption{\label{TrimersFormation} (a) Triangular lattice formed by V ions (shown in red) in the $ab$ plane in the high-temperature ($R\bar3m$) phase of LiVO$_2$ and (b) the resulting trimerized structure in the low-temperature phase of LiVO$_2$ with V trimers (corresponding short V-V bonds are shown in purple thick lines). Thin red lines demonstrate longer V-V bonds.}
\end{figure}

The possible tripling of the unit cell was predicted by theoretical calculations, where the high-temperature phase was found to be unstable~\cite{Sub}. Frequencies of two acoustic modes of the phonon spectrum at  $\bf q = $(1/3,-1/3,0) with respect to the reciprocal lattice vectors of the primitive unit cell turned out to be negative. This led Subedi to suggestion that the primitive cell of the low-temperature phase with three formula units (12 atoms) has a monoclinic $Cm$ structure with the V ions forming trimers shown in Fig.\ref{TrimersFormation}(b). Another version of the low-temperature crystal structure was proposed in ~\cite{Onoda}, and then refined taking into account the pair distribution function (PDF) analysis in~\cite{Pour}. In a very recent PDF study by Kojima {\it et al.} it was shown that apparent difficulties in crystal structure determination in LiVO$_2$ are related with a random stacking of V trimers along $c$ direction~\cite{Kojima2019,Kojima2023}. Thus, even after more than 70 years of investigations the crystal structure of this material remains unsolved and, moreover, there are indications that stacking faults are always present and one should rather discuss presence of a short-range trimer order in the stacking direction~\cite{Kojima2023}. 

This makes studying of LiVO$_2$  {\it local} structure especially important. While Raman spectroscopy is known to be extremely sensitive to local changes of the crystal structure, surprisingly no detailed analysis of Raman data for LiVO$_2$ have been performed so far. In this work, we measured symmetry-dependent Raman spectra of  LiVO$_2$ three single crystals and study their temperature behavior across the phase transition. Experimental results are supplemented by the density functional theory (DFT) calculations of the phonon spectra and possibility of the stacking faults. The results showed the surprising sensitivity of Raman spectroscopy to structural changes in the sample, thereby confirming the formation of superstructures due to trimerization of V ions.


\section{Experimental and calculation details}\label{details}
\begin{figure}[t!]
\centering
\includegraphics[width=1\columnwidth]{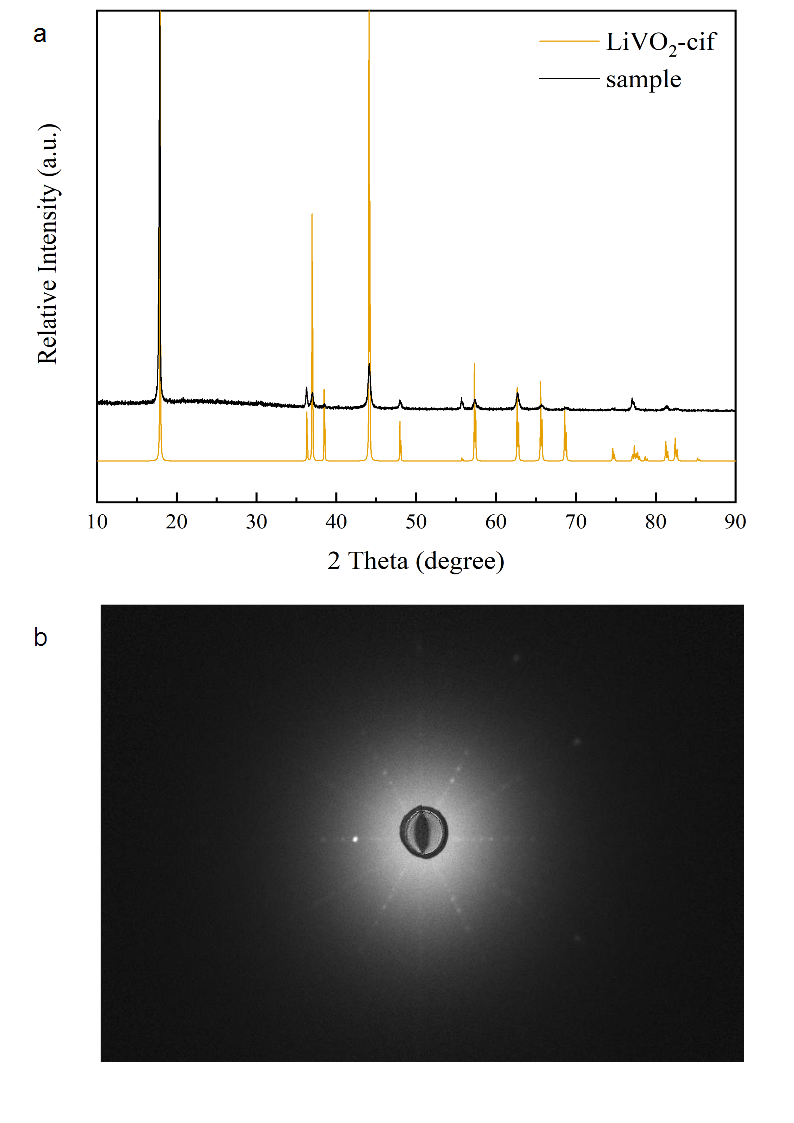}
\caption{\label{XRD} (a) X-ray diffraction (XRD) pattern and (b) Laue picture along the (001) direction of grown LiVO$_2$ single crystal. }
\end{figure}

Small single crystals (of size 2$\times$2$\times$0.5 mm$^3$) of LiVO$_2$ were grown by using the flux method as reported before \cite{Tian}. The conventional solid-state-reaction method was applied to prepare LiVO$_2$ powder with pure phase, in which appropriate proportions of Li$_2$CO$_3$ and V$_2$O$_3$ were thoroughly mixed and then calcined at 625$^{\circ}$C and 750$^{\circ}$C under flowing Ar: H$_2$ = 19: 1 with intermediate grinding. Then, the mixture of LiVO$_2$ polycrystalline, Li$_2$O, and LiBO$_2$ was placed in a platinum crucible in the glove box and sealed in a quartz tube. After heating at 1100$^{\circ}$C for 12 h, the tube was cooled to 700$^{\circ}$C by oscillating process, then turned off the power. The chemical composition of the sample was checked by the Inductively coupled plasma (ICP) mass spectroscopy (NexION 350X), giving the ratio of Li: V = 1.01. The phase purity was identified by the X-ray diffraction (XRD, Aeris, CuK$\alpha$1 radiation) and the X-ray Laue back diffraction was used to confirm the quality of the crystal, see Fig.~\ref{XRD}.

We performed Raman measurements in the temperature range of 80 to 600 K in backscattering geometry.  The measurements were performed with RM1000 Renishaw microspectrometer equipped with a 532 nm solid-state laser and 633 helium–neon laser and with a thermocooled  CCD camera. Respective Linkam stage was used for temperature variation. Very low power (up to 1 mW) was used to avoid local heating of the sample. A pair of notch filters with a cut-off at 40 cm$^{-1}$ were used to reject light from the 633 nm laser line. To reach as close to the zero frequency as possible, we used a set of three volume Bragg gratings (VBG) at 532 nm excitation to analyze the scattered light. The resolution of our Raman spectrometer was estimated to be 2–3 cm$^{-1}$.

All the density functional theory (DFT) calculations were performed in Vienna ab initio simulation package (VASP) \cite{Kresse4}. We utilized generalized gradient approximation (GGA) as proposed by Perdew, Burke, and Ernzerhof (PBE)~\cite{Perdew} and the cutoff energy for the plane-wave basis set is 500 eV. The DFT iterations were repeated until the total energy change was more than 10$^{-8}$ eV between electronic steps and more than 10$^{-6}$ eV between ionic steps.

Calculations of the phonon dispersion curves for the high-temperature phase ($R\bar 3m$) were performed using frozen-phonon technique implemented in Phonopy code \cite{phonopy}. For this purpose the 3$\times$3$\times$3 supercell constructed from the primitive one was used. For such a cell 6$\times$6$\times$6 mesh of Brillouin zone in the reciprocal space was chosen. The experimental crystal structure was fully relaxed before lattice dynamics simulations.

Phonon frequencies for the low-temperature structure (space group $Cm$) were obtained only for the $\Gamma$-point of the Brillouin zone using density functional perturbation theory (DFPT) \cite{dfpt}. For the crystal structure suggested in \cite{Sub} only atomic positions were optimized and 6$\times$6$\times$6 mesh of Brillouin zone in the reciprocal space was used.

Structural optimization for the considered stacking patterns in the low-temperature structure (space groups $Cm$ and $Cc$) were performed with the 8$\times$6$\times$4 mesh of Brillouin zone in the reciprocal space.

\section{Experimental results }
In order to obtain information about  symmetry of the observed excitations, polarization measurements were performed in two geometries, with parallel (XX) and  mutually perpendicular (XY) polarizations of the incident and scattered light, where X and Y refer to the directions in the basal plane of crystals that have a hexagonal shape. In case of $R\bar 3m$ crystal structure XX geometry is characterized by the $A_{1g}$ + $E_g$ symmetry, and XY - by the $E_g$ symmetry. Thus, these two measurements make it possible to obtain information about the behavior of pure symmetric components of the spectrum.
\begin{figure}[b!]
\centering
\includegraphics[width=1\columnwidth]{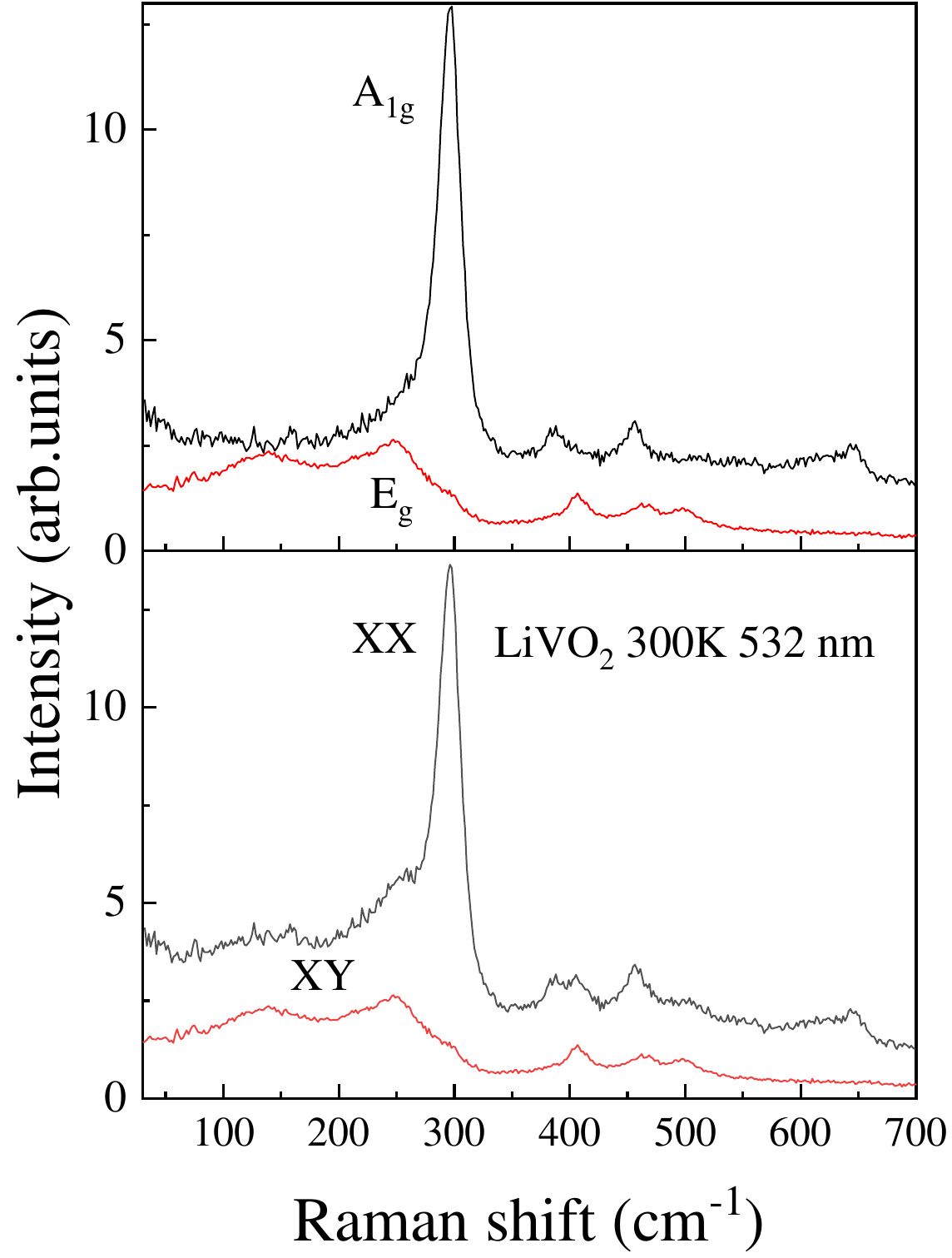}
\caption{\label{RT} Raman spectra of as-grown samples at 300K for both scattering symmetries (low panel) and extracted symmetry-dependent contributions (upper panel). The 532 nm laser was used.}
\end{figure}

The high-temperature (above 500 K) rhombohedral phase  with $R\bar 3m$ space group has the $A_{1g}+2A_{2u}+2E_u+E_g$ optical phonons at the center of the Brillouin zone. Two of them, $E_g$ and $A_{1g}$, are Raman active.  They are observed at 486 and 595 cm$^{-1}$ in LiCoO$_2$ ~\cite{Co}, at 465 and 544 cm$^{-1}$ in  LiNiO$_2$~\cite{Ni}, and at 456 and 589 cm$^{-1}$ in LiCrO$_2$~\cite{Cr}. For the $A_{1g}$ mode, the atomic shift of oxygen atoms is along the $c$-axis, whereas the $E_g$ mode corresponds to vibrations in the basal plane.

\begin{figure}[t!]
\centering
\includegraphics[width=0.7\columnwidth]{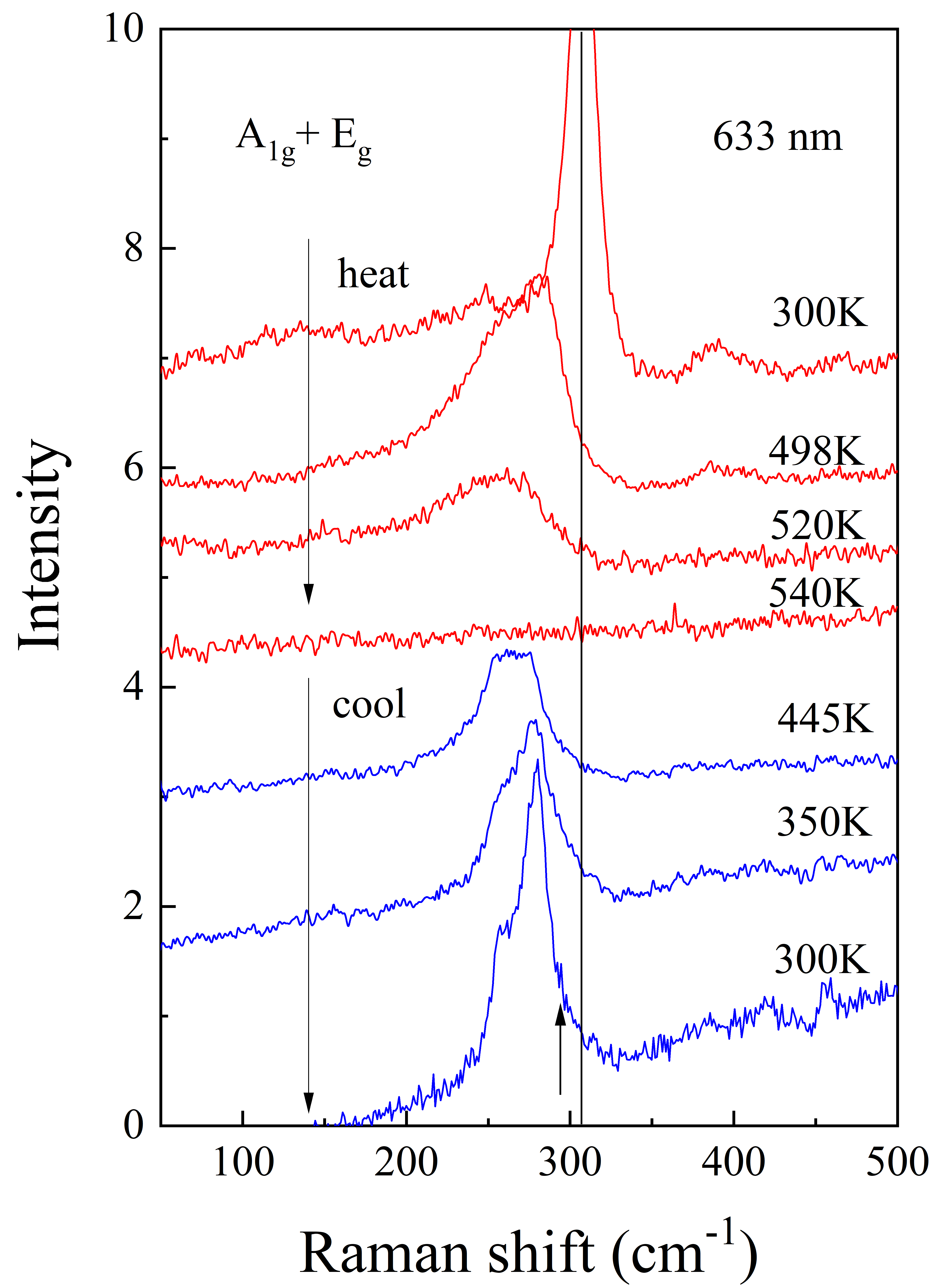}
\caption{\label{1cycling} Temperature dependent Raman spectra in XX scattering geometry measured in the heat-cool cycle of the first sample. Vertical line marks energy position of intensive maximum before cycling.}
\end{figure}

Going through the structural transition our Raman measurements clearly detect lowering of the symmetry in LiVO$_2$. Raman spectra of two measured at 300K LiVO$_2$ samples show a much larger number of phonon lines than those predicted by the selection rules for the $R\bar 3m$ structure (Fig.~\ref{RT}). Three intense lines  are observed in the region of 140-310 cm$^{-1}$.  Moreover, 8-9 low-intensity lines in the XX and 3-4 modes in the XY spectra were measured in the region of 350-700 cm$^{-1}$. Thus, low-frequency broad lines at 144 and 247 cm$^{-1}$, as well as lines at  406, 458, and 495  cm$^{-1}$ are observed in both polarized (XX) and depolarized (XY) geometries (low panel). This is impossible for previously proposed $Cm$ monoclinic structure of the low-temperature phase~\cite{Sub}, where the Raman active modes are $A'$ and $A''$ vibrations. Thus, we can conclude that the symmetry of the low-temperature phase is not lower than trigonal. After subtracting XY from XX spectrum, we obtained pure $A_{1g}$ and $E_g$ spectra (upper panel of Fig.~\ref{RT}) under the assumption of trigonal symmetry of the low-temperature phase. 

A large number of lines observed at low temperatures can be ascribed to change in the lattice symmetry, namely formation of V trimers previously suggested to explain magnetic susceptibility data and recently observed by the PDF analysis~\cite{Kojima2019}. This leads to  a tripling of the unit cell~\cite{Car,Onoda,Tian} and to possible folding of the Brillouin zone. This results in additional active Raman modes in the center of the new Brillouin zone. To confirm this assumption, Raman spectra were measured while temperature cycling through the phase transition point. The close attention will be paid to the low-frequency region of the spectrum, since intense lines in this region are not expected in the Raman spectra for the $R\bar 3m$ structure. 

To sum up this section, the measurements performed clearly detect a radical restructuring of the phonon spectrum as a result of the structural transition. These changes are not only related to the V trimerization in the low-temperature phase, as we will discuss below, but the spectra can be rather sensitive to thermal cycling featuring slow structural dynamics and different degree of the stacking faults in different samples.


\subsection{Thermal measurements}

\begin{figure}[t]
\centering
\includegraphics[width=1\columnwidth]{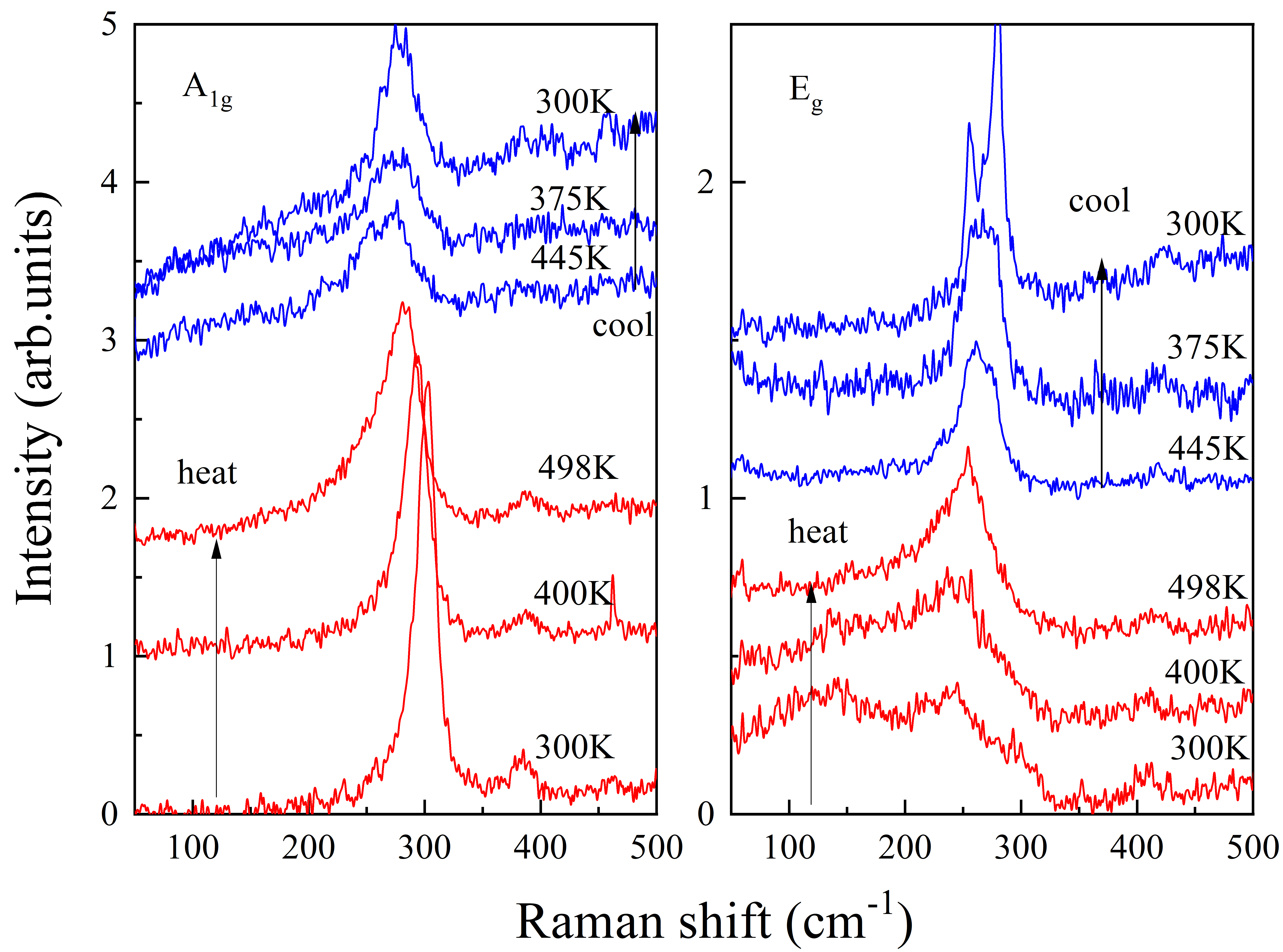}
\caption{\label{1AE-cycling} Evolution of the $A_{1g}$ and $E_g$ Raman spectra at the first temperature cycle.}
\end{figure}

We performed thermal measurements on two seemingly similar single crystals. In Fig.~\ref{1cycling} we present Raman spectra measured at different temperatures in the XX geometry. One can see that after heating the first sample to 540 K, the spectrum measured in this geometry completely disappears.   Upon subsequent cooling down to 445 K it recovers, but the frequency of main peak turns out to be lower than initial. Fig.~\ref{1AE-cycling} illustrates spectral changes under thermal cycling for the both scattering symmetries. Characteristic feature of as-grown sample is the hardening of broad (80-100 cm$^{-1}$) lines near 142 and 245 cm$^{-1}$ in the  $E_g$ spectrum as temperature increases. Finally these two modes almost merge into an asymmetric line at 253 cm$^{-1}$ at 520 K shortly before its disappearance. This behavior of the $E_g$ modes with increasing temperature contrasts to the softening observed for the $A_{1g}$ mode. After cooling to 300 K the $E_g$  spectrum contains two narrow lines at 256 and 279 cm$^{-1}$, which apparently indicates that ordering has occurred. At the same time, the line in the $A_{1g}$ spectrum shows a softening by $\sim$20 cm$^{-1}$ and broadening compared to the original spectrum.

\begin{figure}[b!]
\centering
\includegraphics[width=1\columnwidth]{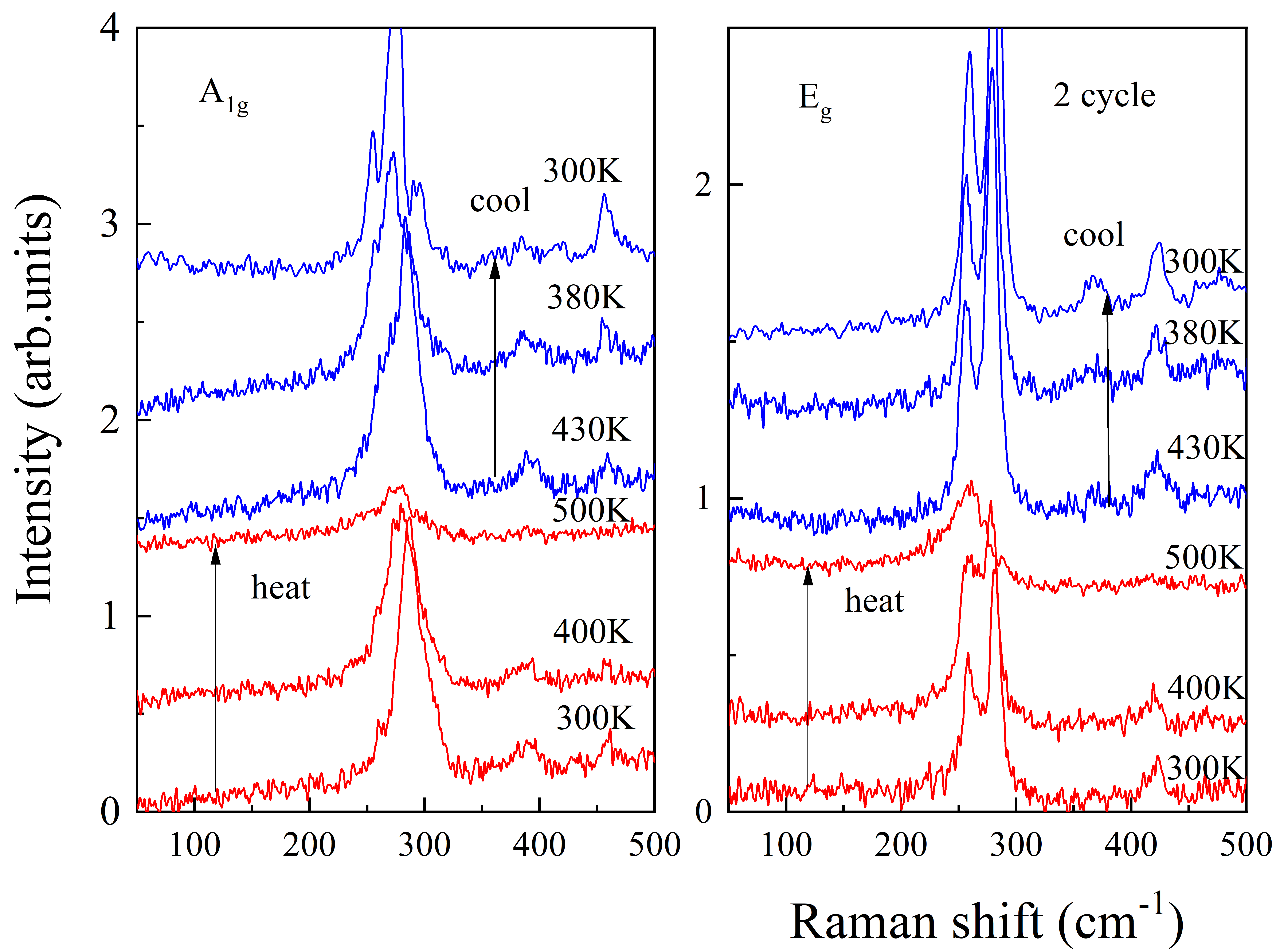}
\caption{\label{2AE-cycling}  $A_{1g}$ and $E_g$ Raman spectra at  the second temperature cycle.}
\end{figure}

These changes in Raman spectra after thermal cycling can evidence the slow structural reconstruction, which likely engage modifications of the stacking order. In order to study this effect further on we performed the second cycle of thermal scanning through the magnetic transition temperature and the spectrum disappeared after 520 K (540 K in the first run) with heating and it recovered below 450 K with cooling (Fig.~\ref{2AE-cycling}). After cooling, the energies of the $E_g$  lines returned to their original values, remaining  narrow, while the frequency of the $A_{1g}$ line decreased and showed the presence of a shoulder or sometimes two shoulders, depending on point of the sample, where the measurement was performed.
\begin{figure}[t!]
\centering
\includegraphics[width=0.8\columnwidth]{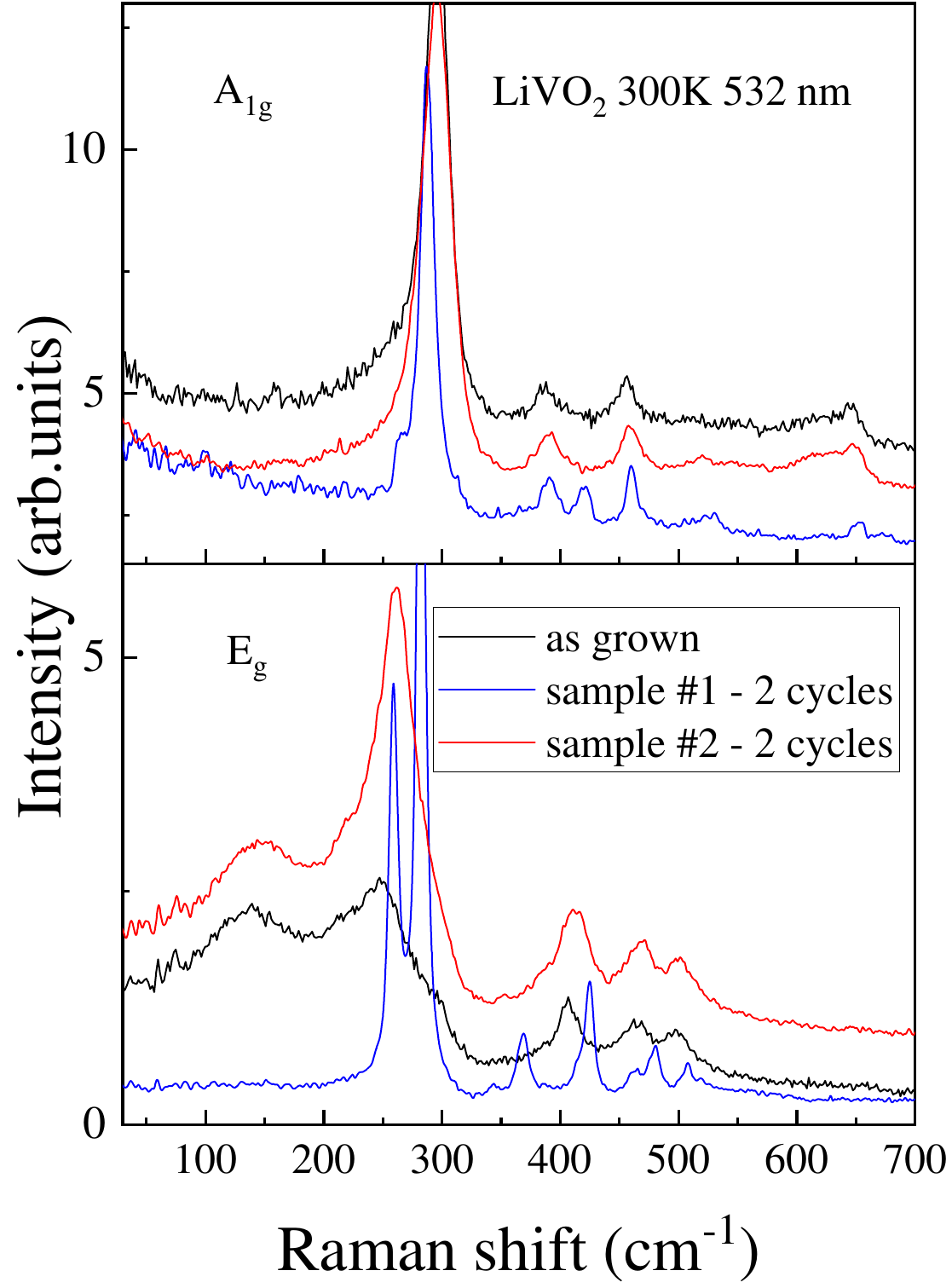}
\caption{\label{12after}  $A_{1g}$ and $E_g$ Raman spectra of the first and second samples before and after 2 thermal cycles.}
\end{figure}

Results obtained during temperature cycling of the second sample are in qualitative agreement with those obtained for the first crystal. The spectra for both symmetries became identical and almost completely disappeared at 510K (Fig. S1)~\cite{Supp}. When passing the temperature of magnetic transition, the sample observed on the video camera showed a rather strong and sharp movement in the basal plane of the crystal. A low-intensity broad band in the region of 210 cm$^{-1}$ persisted at higher temperatures. When the sample was cooled below 465K, broad peaks at 260 cm$^{-1}$ appeared in the spectra of both symmetries, which acquired individual features upon further cooling.

However, the $A_{1g}$ and  $E_g$ spectra measured at 300K were noticeably different from the spectra obtained after thermal cycling on the first sample, as can be seen from Fig. S2 and Fig.~\ref{12after}, which shows the spectra of both as-grown samples before and after repeated thermal cycling. Despite the fact that both samples were synthesized under the same conditions and their initial spectra at 300 K are the same, the spectra of the second sample after thermal cycling are very similar to the spectra of as-grown samples, while the spectra of the first sample show a more ordered structure in both the low- and high-frequency regions. Most likely, these differences are associated with different ordering of the trimers in the direction perpendicular to the \textit{ab} plane. The second case may be due the trimer disorder in the stacking direction while the first one obviously represents a well-ordered structure of trimerized layers in the stacking direction. Although these results were obtained from measurements at various points on both crystals, including both natural and cleaved areas, the reason for this discrepancy is currently unclear. It should be noted that despite the different degrees of ordering of the trimers in the two crystals studied, the observed temperatures of the structural transition are very close.

The temperature behavior of the low-frequency $E_g$ mode near 150 cm$^{-1}$ turned out to be unusual, since its frequency increases when heated up to 500 K in all investigated crystals. But if in the first crystal (supposedly more ordered) frequency of this mode practically did not change during the reverse transition, then in the second  crystal its frequency decreases from ~210 to 144 cm$^{-1}$ upon cooling from 465K to 300K and practically returned to its original position, see Fig. S2. Upon further cooling down to 85 K it softens on another 40 cm$^{-1}$, see Fig. S3 (“soft mode”?). Similar low-temperature results are observed for the third crystal for various phonon symmetries (Fig. S4).  The observed temperature behavior suggests that this soft mode is a folded counteract of one acoustic branch of the $R\bar 3m$ phase, which becomes unstable during the magnetic transition ~\cite{Sub}. However, the softening slows down at low temperatures, suggesting stability of the observed phase. It should also be noted that this behavior is observed only in disordered sample (sample 2), which indicates better stability of the first crystal with ordered trimerized layers.

As for the $A_{1g}$ and $E_g$ phonons allowed for the $R\bar 3m$ structure at temperatures above the magnetic transition point, we observed a broad line of the A$_{1g}$ symmetry near 580 cm$^{-1}$, which significantly increases (up to 630 cm$^{-1}$) upon cooling below the transition temperature. Such strong changes in a rather narrow temperature range of ~30 K are probably associated with a strong decrease of the lattice constant \textit{a} ~\cite{Tian}. A broad band (100 cm$^{-1}$) of the $E_g$ symmetry is observed at high temperatures near 430 cm$^{-1}$, but it is difficult to attribute it definitely to any of the three narrow $E_g$ lines that develop in this energy range when the crystal is cooled below the transition temperature.

To sum up this section, the thermal cycling can reveal difference in two seemingly similar LiVO$_2$ single crystals, which is likely due to different degree of stacking faults (our first sample is more ordered that the second one). 

\subsection{Analysis of low-temperature data~\label{theory}}
\begin{figure}[b!]
\centering
\includegraphics[width=0.7\columnwidth]{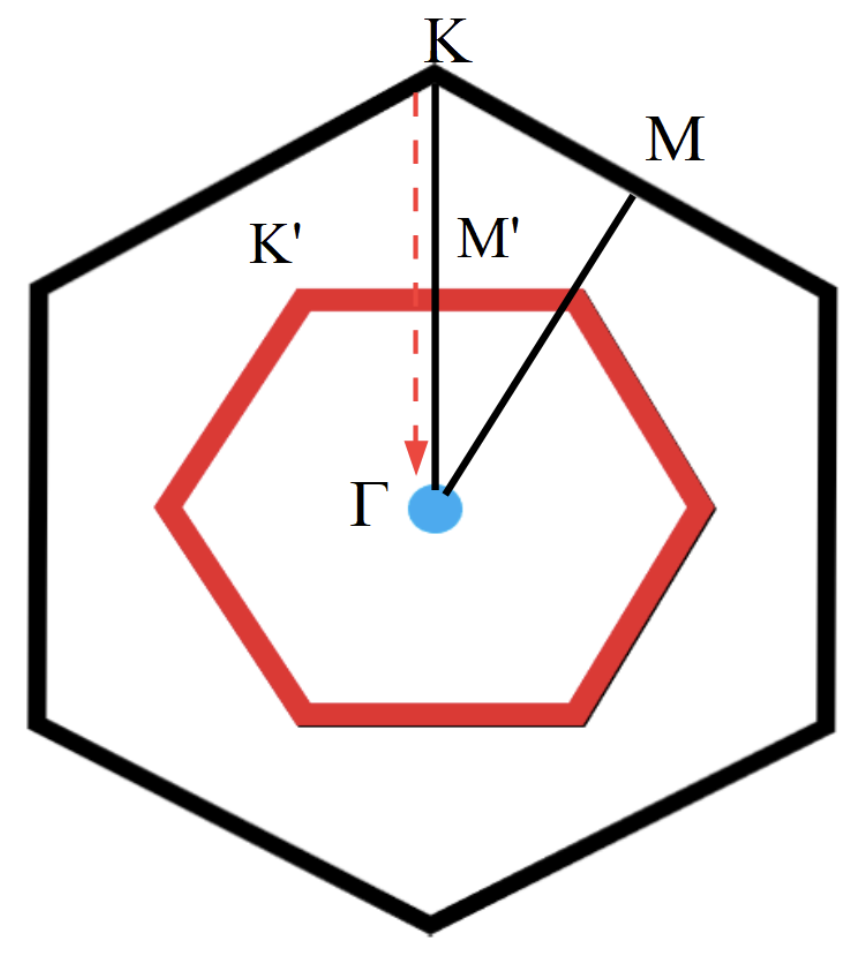}
\caption{\label{Bril}  Reciprocal space of the triangular lattice with black and red hexagons denoting the high-temperature ($R\bar 3m$) and possible low-temperature Brillouin zones.}
\end{figure}

The proposed cell of the low-temperature phase ~\cite{Onoda,Pour} has a lattice parameter $a^* = \sqrt{3} a$ and is rotated by 30$^{\circ}$ relative to the initial (high-temperature $R\bar 3m$) one. 
 The wave vector (1/3,-1/3,0)  for the primitive $R \bar 3m$ cell (in notations of the equivalent hexagonal Brillouin zone (BZ)-the point K) is exactly 2 times larger than the wave vector of the $M'$ point  on the boundary of the reduced BZ (Fig.~\ref{Bril}). Consequently, as a result of the BZ folding, the phonons from  initial zone boundary at the $K$ point  may appear in the center of the new zone and become Raman active. Thus, it can be assumed that 3 intense low-frequency lines are the result of the folding of the acoustic $A_{2u}$ and $E_u$ branches in the new BZ, and high-frequency lines are result of the folding of the $A_{1g}$, $A_{2u}$, $E_g$, and $E_u$ optical branches. The experiment suggests that each of the $E_u$ acoustic branches forms the $E_g$ doublet at the $\Gamma$ point, so similar symmetry is expected when folding the 2 $E_u$ optical modes. In all measurements, we observed an asymmetric $A_{1g}$ line (or two lines) at 280-300 cm$^{-1}$, so it is natural to assume that the $A_u$ branches fold up into $A_{1g}$ at the $\Gamma$ point.
 In such a scenario together with the original optical modes at $\Gamma$ the phonon spectrum of the low-temperature phase will contain at least 33 optical modes. 
 
\begin{figure}[t]
\centering
\includegraphics[width=1\columnwidth]{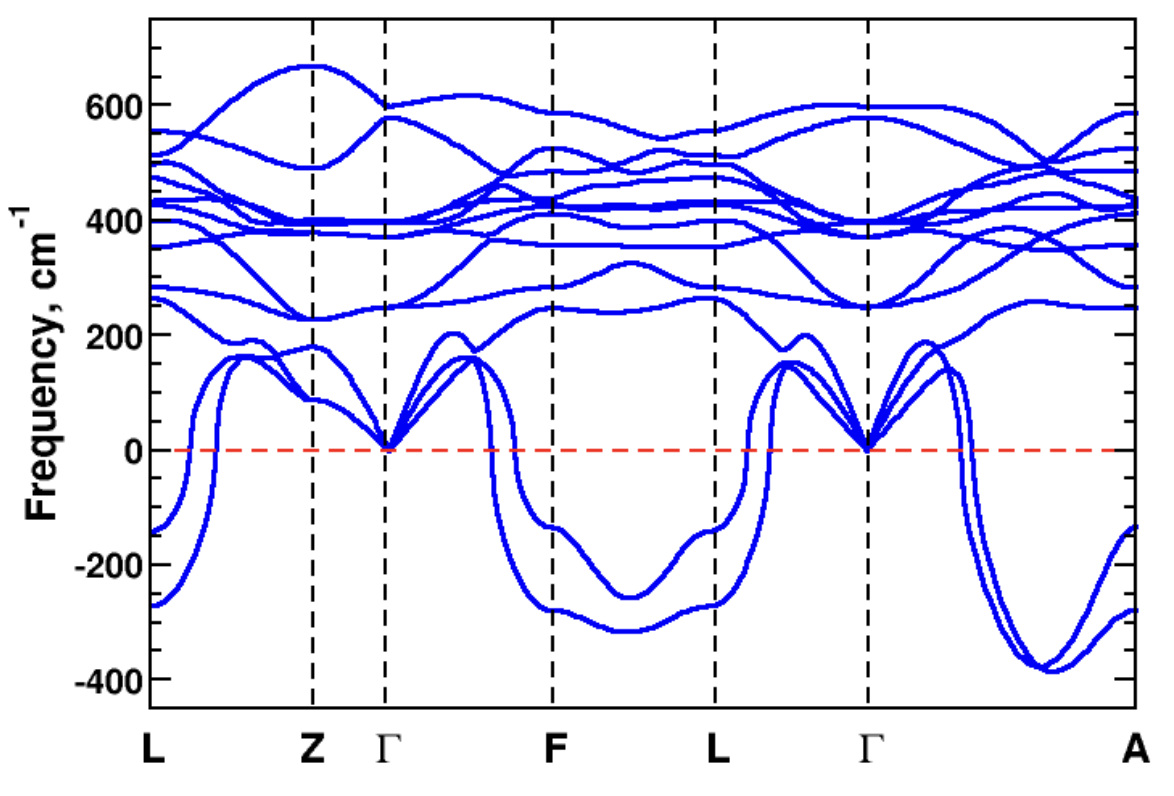}
\caption{\label{Spectrum-HT} Calculated in non-magnetic GGA phonon dispersion curves for the high-temperature ($R\bar 3m$) phase of LiVO$_2$. High-symmetry points of the Brilluoin zone are taken for a primitive cell in rhombohedral axis. Absolute minimum in the negative values in $\Gamma$-A direction corresponds to the $\bf{q}$-vector $(\frac{1}{3}, -\frac{1}{3},0)$ (further in the text the corresponding point is named K after the point in cell with the hexagonal axis).}
\end{figure}

As can be seen from Fig.~\ref{12after}, at least 16 lines (8-9 A$_{1g}$ and 8-9 E$_g$) are observed in the spectra and other may be nonactive in Raman. To confirm the fact of phonon back-folding we tried to compare the obtained spectra with our calculations of the phonon spectrum of the high-temperature phase. For this comparison, we used the calculated for $R\bar 3m$ phase  frequencies at the $\Gamma$-A direction. (Fig.~\ref{Spectrum-HT}). 

The comparison presented in Tab.~\ref{freq} shows some agreement (especially in the high-frequency range) between the experimental and calculated frequencies, which provides arguments in favor of observing folded phonons. The result is quite striking, since the observation of folded phonons in a Raman spectrum is an uncommon phenomenon. Such spectra are well known in the case of measurements of different superlattices (e.g. various SiC polytypes ~\cite{SiC} or layered systems having charge density instabilities ~\cite{Alb,Sam}); sometimes new lines were interpreted as phonons activated by the appearance of the magnetic Brillouin zone as a result of a magnetic transition~\cite{Bal}.

Although the exact crystal structure of the trimerized phase is unknown (most probably because of the stacking faults) we compare the experimental and calculated optical frequencies for the $Cm$ structure~\cite{Sub} (they are presented in the last column of Tab.~\ref{freq}). Although experimental phonon symmetries do not agree with those obtained in these calculations, the frequency range of the observed and calculated energies agrees quite well.


\begin{table}[t]
 \begin{tabular}{c c c c c c}  
  \hline \hline
\multicolumn{3}{c}{Experiment} & \multicolumn{3}{c}{Calculations} \\

\multirow{ 2}{*}{No}. & \multirow{ 2}{*}{Symmetry} & \multirow{ 2}{*}{Mode} & \multicolumn{2}{c}{$R\bar3m$} & $Cm$ \\
    &          &       & $\Gamma$-point & K-point   & $\Gamma$-point\\
 \hline
 1 & $E_g$    & 260 &                   &          & 263 \\
 2 & $A_{1g}$ & 268 & 249               & 258      & 265 \\
 3 & $E_g$    & 281 &                   &          & 277 \\
 4 & $A_{1g}$ & 287 &                   &          & 279 \\
 5 & $E_g$    & 343 &                   & 349      & 326, 327, 333, 349 \\
 6 & $E_g$    & 368 & \textbf{372}      & 371, 374 & 372, 377 \\
 7 & $A_{1g}$ & 390 & 397               &          & 382, 394, 397 \\
 8 & $A_{1g}$ & 420 & 398               & 421      & 409, 410 \\
 9 & $E_g$    & 424 &                   &          & 435 \\
10 & $A_{1g}$ & 460 &                   & 447      & 447, 456 \\
11 & $E_g$    & 462 &                   &          & 463, 464 \\
12 & $E_g$    & 480 &                   & 483      & 480, 481 \\
13 & $E_g$    & 507 &                   & 495, 500 & 509 \\
14 & $E_g$    & 520 &                   & 505      & 526 \\
15 & $A_{1g}$ & 527 & 573               &          & 532 \\
16 & $A_{1g}$ & 618 & 598, \textbf{599} &          & 592, 596 \\
17 & $A_{1g}$ & 650 &                   &          & 665, 704 \\
\hline
\hline
 \end{tabular}
	\caption{\label{freq}Comparison between experimentally found (for the first sample after thermal cycling) and calculated for both high-temperature (at $\Gamma$ and at K$=(\frac{1}{3},-\frac{1}{3},0)$) and low-temperature (at $\Gamma$) phases 
	phonon frequencies 
 given in cm$^{-1}$. Symmetry are given for the experimental modes. The calculated Raman active modes in the $\Gamma$-point for the $R\bar3m$ structure are given in bold.}
		\label{table}
\end{table}


\section{Calculations of different types of stacking}

\begin{figure}[b!]
\centering
\includegraphics[width=1\columnwidth]{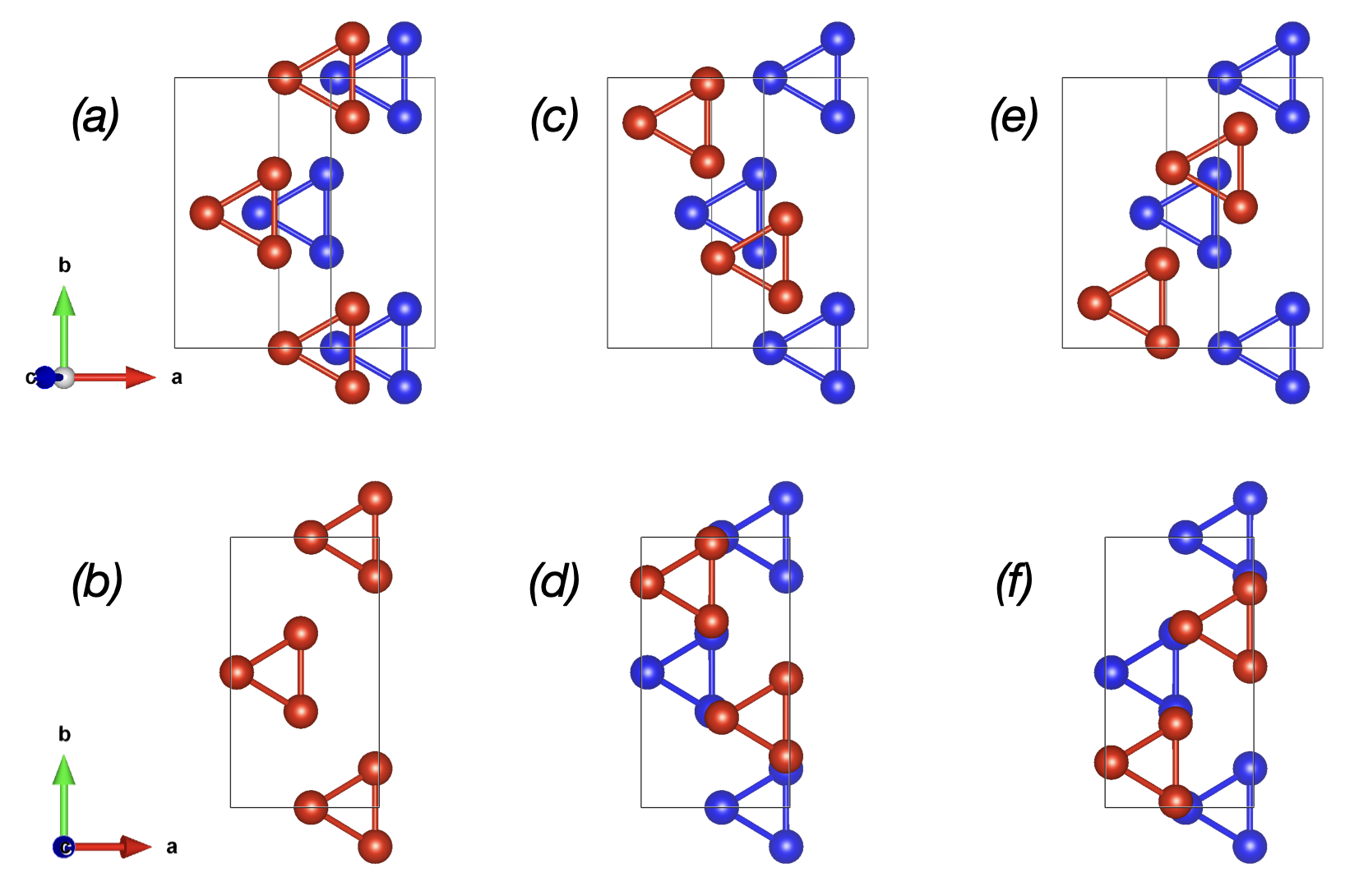}
\caption{\label{stacking} Considered types of stacking in LiVO$_2$ along the $c$-axis. Only vanadium ions in the doubled primitive cell are shown. V in the nearest layers are given in different colors (blue for first layer in the cell and red for the upper one). (a-b) Initial structure suggested in \cite{Sub} (space group $Cm$), (c-d) the structure with the $b/3$ and (e-f) $2b/3$ shift for the nearest V layer (in comparison to the initial one).}
\end{figure}

In order to check the possibility of stacking faults in LiVO$_2$ one can estimate the total energy difference between the types of stacking. We started with the structure suggested by Subedi in \cite{Sub} (Fig.~\ref{stacking}(a-b)) and doubled the primitive cell along the $c$-axis in order to introduce corresponding atomic shifts. The lower-lying layer of V atoms was untouched while the upper one was shifted by either $b/3$ (Fig.~\ref{stacking}(c-d)), or $2b/3$ (Fig.~\ref{stacking}(e-f)). This immediately changes symmetry of the structure to the $Cc$ space group.

We performed structural optimization (in non-magnetic GGA) for all three stacking types and all the initial patterns were retained in the resulting structures demonstrating their stability. The total energies of different stacking types differ by a tiny amount of 0.1 meV/f.u. Thus, one could naturally expect stacking faults in LiVO$_2$, which explains not only previous X-ray experiments, but also our Raman measurements, when two seemingly similar samples show a different spectra under thermal cycling.

\section{\label{sec:concl}Conclusions}

In the present paper we demonstrate that Raman spectroscopy is a sensitive probe helpful in studying a long-standing problem of describing the structural and magnetic transition observed in LiVO$_2$ at $\sim$ 500K. It clearly shows that the symmetry of a low-temperature phase supposedly having trimerized crystal structure (V trimers) is not lower than trigonal. Detailed analysis of Raman spectra of single crystals under thermal cycling evidence a slow structural dynamics (e.g. slow rearranging of V trimers) or stacking faults. The degree of stacking faults can be different in seemingly similar samples. Thus, the Raman spectroscopy can be used as a hallmark for testing LiVO$_2$ samples in further studies. Theoretical calculations based on the density functional theory supports possible stacking faults demonstrating that the energy of different stacking types is of order 1 K per formula unit.

\section{Acknowledgments}\normalfont
The authors are grateful to A.V. Ushakov and R. Maezono for useful discussions and I.V. Baklanova for help with low-temperature measurements.  The research was carried out within the state assignment of Ministry of Science and Higher Education of the Russian Federation (themes “Electron” No 122021000039-4 and ``Quantum'' No 122021000038-7). A portion of this work was supported by National Key R\&D Program of China (No. 2023YFA1406100).

\end{document}